\ifdefined\isabridged  
\documentclass{llncs}
\else                  
\documentclass[sigconf]{acmart}
\fi
\usepackage[T1]{fontenc}
\usepackage[utf8]{inputenc}
\usepackage{letltxmacro}
\usepackage{todonotes}
\usepackage{xspace}
\usepackage{hyperref}
\usepackage{pifont}
\usepackage{multirow}
\usepackage{tablefootnote}
\usepackage{float}
\usepackage{listings}
\usepackage{times}

\newif\ifabridged
\newif\ifnotabridged
\newif\ifanonymous
\newif\ifnotanonymous
\newif\ifllncs
\newif\ifacmart

\ifdefined\isabridged
\abridgedtrue
\fi

\ifabridged\notabridgedfalse
\else\notabridgedtrue
\acmarttrue
\fi

\ifdefined\isanonymous
\anonymoustrue
\fi

\ifanonymous\notanonymousfalse
\else\notanonymoustrue
\fi

\makeatletter  
\@ifclassloaded{llncs}{\llncstrue}{\llncsfalse}
\@ifclassloaded{acmart}{\acmarttrue}{\acmartfalse}
\makeatother

\LetLtxMacro{\todonote}{\todo}
\renewcommand{\todo}[2][]
{\todonote[inline, caption={#2}, size=\footnotesize, #1]
{\renewcommand{\baselinestretch}{0.5}\selectfont#2\par}}

\renewcommand\appendix{\par
\setcounter{section}{0}
\setcounter{subsection}{0}
\renewcommand\thesection{APPENDIX \Alph{section}} }

\ifllncs       
\makeatletter  
\renewcommand\section{\@startsection{section}{1}{\z@}%
  {-8\p@ \@plus -4\p@ \@minus -4\p@}%
  {6\p@ \@plus 4\p@ \@minus 4\p@}%
  {\normalfont\large\bfseries\boldmath
\rightskip=\z@ \@plus 8em\pretolerance=10000 }}
\renewcommand\subsection{\@startsection{subsection}{2}{\z@}%
  {-8\p@ \@plus -4\p@ \@minus -4\p@}%
  {6\p@ \@plus 4\p@ \@minus 4\p@}%
  {\normalfont\normalsize\bfseries\boldmath
\rightskip=\z@ \@plus 8em\pretolerance=10000 }}
\renewcommand\subsubsection{\@startsection{subsubsection}{3}{\z@}%
  {-4\p@ \@plus -4\p@ \@minus -4\p@}%
  {-1.5em \@plus -0.22em \@minus -0.1em}%
{\normalfont\normalsize\bfseries\boldmath}}
\makeatother

\def\baselinestretch{0.95}

\ifnotanonymous  
\author{Thomas Nyman\inst{1}\textsuperscript{,}\inst{2}
  \and Jan-Erik Ekberg\inst{2}
  \and Lucas Davi\inst{3}
  \and N. Asokan\inst{1}
}

\institute{Aalto University, Finland \\
  \email{\normalsize thomas.nyman@aalto.fi},\email{\normalsize n.asokan@acm.org}
  \and Trustonic, Finland \\
  \email{\normalsize \{thomas.nyman,jee\}@trustonic.com}
  \and University of Duisburg-Essen, Germany \\
  \email{\normalsize lucas.davi@wiwinf.uni-due.de}
}
\fi           
\fi           

\ifacmart     

\makeatletter 
\def\@typeset@author@bx{\bgroup\hsize=.2\textwidth\relax\def\and{\par}%
  \global\setbox\author@bx=\vtop{\if@ACM@sigchiamode\else\centering\fi
  \@authorfont\@currentauthors\par\@affiliationfont
  \@currentaffiliation}\egroup
  \box\author@bx\hspace{\author@bx@sep}%
  \gdef\@currentauthors{}%
  \gdef\@currentaffiliation{}}
\makeatother

\ifnotabridged 
\renewcommand\footnotetextcopyrightpermission[1]{} 
\fi

\begin{abstract}
With the increasing scale of deployment of Internet of Things (IoT), concerns about IoT security have become more urgent. In particular, memory corruption attacks play a predominant role as they allow remote compromise of IoT devices. Control-flow integrity (CFI) is a promising and generic defense technique against these attacks. However, given the nature of IoT deployments, existing protection mechanisms for traditional computing environments (including CFI) need to be adapted to the IoT setting. In this paper, we describe the challenges of enabling CFI on microcontroller (MCU) based IoT devices. 
We then present \tzmcfi, the first \emph{interrupt-aware} CFI scheme for low-end MCUs. \tzmcfi uses a novel way of protecting the CFI metadata by leveraging TrustZone-M security extensions introduced in the ARMv8-M architecture. Its binary instrumentation approach \emph{preserves the memory layout} of the target MCU software, allowing pre-built bare-metal binary code to be protected by \tzmcfi. We describe our implementation on a Cortex-M Prototyping System and demonstrate that \tzmcfi is secure while imposing acceptable performance and memory impact.
\end{abstract}

\ifnotanonymous  
\author{Thomas Nyman}
\affiliation{
  \institution{Aalto University, Finland}
}
\email{thomas.nyman@aalto.fi}
\affiliation{
  \institution{Trustonic, Finland}
}
\email{thomas.nyman@trustonic.com}

\author{Jan-Erik Ekberg}
\affiliation{
  \institution{Trustonic, Finland}
}
\email{jee@trustonic.com}

\author{Lucas Davi}
\affiliation{
  \institution{University of Duisburg-Essen, Germany}
}
\email{lucas.davi@wiwinf.uni-due.de}

\author{N. Asokan}
\affiliation{
  \institution{Aalto University, Finland}
}
\email{asokan@acm.org}
\fi          
\fi          

\DeclareRobustCommand\sectt[1]{{\fontsize{13}{12}#1}}
\newcommand{\tzmcfisectitle}{\protect\sectt{CFI CaRE}\xspace}
\newcommand{\tzmcfi}{\protect{CaRE}\xspace}

\newcommand{\monitorsectitle}{\protect\sectt{Branch Monitor}}
\newcommand{\monitor}{\protect{Branch Monitor}\xspace}

\newcommand{\dOne}{\ding{182}}
\newcommand{\dTwo}{\ding{183}}
\newcommand{\dThree}{\ding{184}}
\newcommand{\dFour}{\ding{185}}
\newcommand{\dFive}{\ding{186}}
\newcommand{\dSix}{\ding{187}}
\newcommand{\dSeven}{\ding{188}}
\newcommand{\dEight}{\ding{189}}

\ifacmart
\newtheorem*{claim}{Claim}
\fi
\newtheorem{requirement}{Requirement}
\newtheorem{assumption}{Assumption}
\newtheorem{observation}{Observation}
\newtheorem{invariant}{Invariant}

\title{CFI CaRE: Hardware-supported Call and Return Enforcement for Commercial Microcontrollers}

\begin{document}
\maketitle

\begin{abstract}
With the increasing scale of deployment of Internet of Things (IoT), concerns about IoT security have become more urgent. In particular, memory corruption attacks play a predominant role as they allow remote compromise of IoT devices. Control-flow integrity (CFI) is a promising and generic defense technique against these attacks. However, given the nature of IoT deployments, existing protection mechanisms for traditional computing environments (including CFI) need to be adapted to the IoT setting. In this paper, we describe the challenges of enabling CFI on microcontroller (MCU) based IoT devices. 
We then present \tzmcfi, the first \emph{interrupt-aware} CFI scheme for low-end MCUs. \tzmcfi uses a novel way of protecting the CFI metadata by leveraging TrustZone-M security extensions introduced in the ARMv8-M architecture. Its binary instrumentation approach \emph{preserves the memory layout} of the target MCU software, allowing pre-built bare-metal binary code to be protected by \tzmcfi. We describe our implementation on a Cortex-M Prototyping System and demonstrate that \tzmcfi is secure while imposing acceptable performance and memory impact.
\end{abstract}

\section{Introduction}
\label{sec:introduction}

\emph{Cyber-Physical Systems} (CPS) are becoming pervasive across many
application areas ranging from industrial applications
(manufacturing), transport, and smart cities to consumer products.
\emph{Internet of Things} (IoT) refers to systems incorporating such
devices with (typically always-on) communication capability.
Estimates put the number of deployed IoT
devices at 28 billion by 2021~\cite{Ericsson15}. 
Although programmable CPS devices are not new, connectivity makes them
targets for network originating attacks.
Gartner highlights device identity (management), code/data integrity and secure communication as the most important security services
for IoT~\cite{Gartner15}.


The system software in IoT devices is often written in memory-unsafe
languages like C~\cite{IoTDeveloperSurvey17}. The arms race~\cite{Szekeres13} in runtime exploitation of general
purpose computers and network equipment has shown us that memory
errors, such as buffer overflows and use-after-free errors, constitute
a dominant attack vector for stealing sensitive data or gaining
control of a remote system. Over the years, a number of platform
security techniques to resist such attacks have been developed and
deployed on PCs, servers and mobile devices. These include protections
against code injection and code-reuse attacks, such as \emph{Control-Flow
  Integrity}~\cite{Abadi09} (CFI) and \emph{Address Space Layout
  Randomization}~\cite{Cohen93,Larsen14} (ASLR) which aim to ensure the
\emph{run-time integrity} of a device.

CFI (Section~\ref{sec:rop}) is a well-explored technique for resisting
the code-reuse attacks such as \emph{Return-Oriented Programming} (ROP)~\cite{Shacham07} that allow 
attackers in control of data memory to subvert the control flow of a 
program. CFI commonly takes the form of inlined enforcement, where CFI
checks are inserted at points in the program code where control flow
changes occur. For legacy applications CFI checks must be introduced by
instrumenting the pre-built binary. Such binary instrumentation
necessarily modifies the memory layout of the code, requiring memory
addresses referenced by the program to be adjusted accordingly~\cite{Habibi15}. This
is typically done through load-time dynamic binary rewriting software~\cite{MSEMET,Davi12}.

A prominent class of state-of-the-art CFI schemes is based on the
notion of a \emph{shadow stack}~\cite{Dang15}: a mechanism that prevents overwriting
subroutine return addresses on the call stack by comparing each
return address to a protected copy kept in the shadow stack before
performing the return. This effectively mitigates return-oriented programming 
attacks that stitch together instruction sequences ending in return instructions~\cite{Shacham07}. 
However, it presumes the existence of mechanisms to ensure that the shadow stack cannot be 
manipulated by the attacker.

As we argue in detail in Section~\ref{sec:cfi}, the type of IoT scenarios we consider have a number of characteristics
that make traditional CFI mechanisms difficult to apply. First, IoT
devices are typically architected as interrupt-driven reactive
systems, often implemented as bare-metal software involving no loading
or relocation.  To the best of our knowledge, no existing CFI scheme is interrupt-aware. 
Second, IoT devices are often based on computing cores that
are low-cost, low-power single-purpose programmable microcontrollers
(MCUs). Countermeasures for general purpose computing devices, such
as ASLR, often rely on hardware features (e.g., virtual memory) that are unavailable in simple MCUs. 
Prior CFI schemes for embedded systems, such as HAFIX~\cite{Davi15}, and the recently announced 
Intel Control-flow Enforcement Technology (CET)~\cite{Intel-CET},
require changes to the hardware and toolchain, access to source code
and do not support interrupts. 

On the positive side, hardware-based
isolation mechanisms for MCUs have appeared not only in the research
literature~\cite{Defrawy12,Koeberl14,Clercq16}, but also as commercial
offerings such as the recently announced TrustZone-M security
extensions for the next generation of ARM microcontrollers
(Section~\ref{sec:tzm}) providing a lightweight trust anchor for
resource-constrained IoT devices~\cite{ARMv8-M}. However, since software (and
updates) for IoT devices may come from a different source than the
original equipment manufacturer (OEM), it is unrealistic to expect the
software vendors to take responsibility for the instrumentation
necessary for hardware-assisted CFI protection -- OEMs in turn will be
incentivized to distribute the same software to \emph{all} devices,
with and without hardware security extensions.



\paragraph{Goal and Contributions.}
\noindent
We introduce the first hardware software co-design based security architecture that (i)~enables 
practical enforcement of control-flow policies, (ii)~addresses the unique challenges of low-end IoT devices with 
respect to CFI deployment, (iii)~requires no changes to the underlying hardware, and (iv)~operates directly on binary 
code thereby avoiding the need for source code. Specifically, we target control-flow integrity policies that 
defend against runtime attacks, such as ROP, that 
belong to the most prominent software attacks on all modern computing architectures, e.g., 
Desktop PCs~\cite{Shacham07}, mobile devices~\cite{Kornau09}, and embedded systems~\cite{Francillon08}. 

To this end we present the design and implementation of a novel architecture, 
\tzmcfi (Call and Return Enforcement), accompanied with a toolchain for achieving robust run-time code integrity for IoT devices. 
We claim the following contributions:
\begin{itemize}
\itemsep0em
  \item The first \textbf{interrupt-aware} CFI scheme for low-end MCUs
    (Section~\ref{sec:tzmcfi}) supporting
    \begin{itemize}
      \itemsep0em
    \item \textbf{hardware-based shadow stack protection} by leveraging recently introduced TrustZone-M security extensions
      (Section~\ref{sec:architecture}).
    \item a new binary instrumentation technique that is memory \textbf{layout-preserving} and can
      be realized \textbf{on-device}
      (Section~\ref{sec:instrumentation}).
    \end{itemize}
  
%

  \item An implementation of \tzmcfi on ARM Versatile Express
    Cortex-M Prototyping System (Section~\ref{sec:implementation}).

  \item A comprehensive evaluation (Section~\ref{sec:evaluation})
    showing that \tzmcfi ensures CFI (Section~\ref{subsec:security}),
    has a lower performance overhead
    (Section~\ref{subsec:performance}) compared to software-based
    shadow stack schemes while imposing comparable impact on program
    binary size (Section~\ref{subsec:codesize}).
\end{itemize}

\section{Background}
\label{sec:background}

\subsection{ARM Architecture} \label{sec:thumb}
\noindent

ARM microprocessors are RISC-based computer designs that are widely
used in computing systems which benefit from reduced cost, heat, and power
consumption compared to processors found in personal computers.
\ifabridged
The ARM Cortex-M series of processors, geared towards low-cost embedded
microcontrollers (MCUs), consists of core designs optimized for different
purposes, such as small silicon footprint (M0), high energy efficiency (M0+),
configurability (M3) or high performance (M4, M7).  Cortex-M processors only
support the 16-bit Thumb and mixed 16 and 32-bit Thumb-2 instruction sets.
32-bit Thumb-2 instructions are encoded as two 16-bit half-words. 
\emph{ARMv8-M}~\cite{ARMv8-M} is the next generation instruction set
architecture for M-class processors. The Cortex-M23\footnote{\url{https://www.arm.com/products/processors/cortex-m/cortex-m23-processor.php}}
and Cortex-M33\footnote{\url{https://www.arm.com/products/processors/cortex-m/cortex-m33-processor.php}}
are the first cores to support the ARMv8-M architecture. Both are compatible
with other processors in the Cortex-M family, allowing (legacy) software re-use
on these devices.
\else 
The ARM architecture is instantiated in three different classes of
processors;

\noindent\textbf{Cortex-A} series application processors are deployed
in mobile devices, networking equipment and other home and consumer
devices.

\noindent\textbf{Cortex-R} series real-time processors are deployed in
embedded devices with strict real-time, fault tolerance and
availability requirements such as wireless baseband processors, mass
storage controllers and safety critical automotive, medical and
industrial systems.

\noindent\textbf{Cortex-M} series of processors are geared towards
embedded microcontrollers (MCUs) requiring minimal cost and high
energy-efficiency such as sensors, wearables and robotics. Some
Cortex-M chips integrate Digital Signal Processing (DSP) and
accelerated floating point processing capability for improved power
efficiency in digital signal control applications.

The current family of Cortex-M processors features five distinct CPUs, ranging
from the low power and energy efficient M0 and M0+ (based on the ARMv6-M
architecture) to the higher performance M3, M4 and M7 CPUs (based on the ARMv7-M
architecture). All these CPUs utilize a mixed 16-bit / 32-bit instruction set
(see \emph{Thumb instruction set} below) and use 32-bit addressing exclusively.
The latest generation of the M-class processor architectures,
ARMv8-M~\cite{Yiu15,ARMv8-M}, is further divided into two subprofiles; \textbf{ARMv8-M
Baseline} for processor designs with a low gate count and a simpler instruction
set (replacing the ARMv6-M architecture), and \textbf{ARMv8-M Mainline} for high
performance embedded systems that demand complex data processing (replacing the
ARMv7-M architecture).

Table~\ref{tbl:devices} shows examples of ARM-based IoT devices together with
their respective processor and memory specifications. In this paper, we focus
primarily on devices in the low end of this device spectrum, i.e., constrained
MCUs (up-to Cortex-M7 equivalent) in single-purpose IoT devices, where memory
and storage are at a premium. The ARM9, ARM11 and Cortex-A -based devices in
Table~\ref{tbl:devices} typically run general purpose operating systems where
the kernel security architecture can employ numerous access control and
enforcement mechanisms, such as memory isolation and access control on virtual
system resources. In contrast the devices we focus on lack a \emph{Memory
Management Unit} (MMU) and the role of the \emph{Operating System} (OS) is
reduced to basic scheduling tasks. Rudimentary isolation capabilities between
distinct software components may be provided through the presence of a
\emph{Memory Protection Unit} (MPU), in which case the configuration of the MPU
is tasked to the OS.
\fi

\ifnotabridged
\begin{table*}[!tb]
\begin{center}
    \caption{Comparison between some contemporary 32-bit ARM IoT-related devices}
    \label{tbl:devices}
    \begin{tabular}{ | l | r | r | r | l l |}
    \hline
    \multicolumn{1}{|c|}{\textbf{Device}} &
    \multicolumn{1}{|c|}{\textbf{Cores $\times$ Clock}}  &
    \multicolumn{1}{|c|}{\textbf{RAM}}    &
    \multicolumn{1}{|c|}{\textbf{Flash}}  &
    \multicolumn{2}{|c|}{\textbf{Processor}} \\
    \hline
    Kionix Sensor Hub
    \tablefootnote{\url{http://www.kionix.com/product/KX23H-1035}}
    & 32 MHz   & 16 KB & 128 KB  & KX23H-1035 & (ARM Cortex-M0) \\
    \hline
    Polar V800 watch
    \tablefootnote{\url{https://community.arm.com/groups/wearables/blog/2015/01/05/a-closer-look-at-some-of-the-latest-arm-based-wearables}}
    & 180 MHz  & 256 KB & 2 MB   & STM 32F437 & (ARM Cortex-M4) \\
    \hline
    Atmel in-vehicle entertainment
    \tablefootnote{\url{http://www.atmel.com/applications/automotive/infotainment-hmi-connectivity/audio-amplifier.aspx}}
    & \multirow{2}{*}{300 MHz}
    & \multirow{2}{*}{384 KB}
    & \multirow{2}{*}{2 MB}
    & \multirow{2}{*}{ATSAMV71Q21}
    & \multirow{2}{*}{(ARM Cortex-M7)} \\
    remote audio amplifier & & & & & \\
    \hline
    Nintento 3DS
    \tablefootnote{\url{https://3dbrew.org/wiki/Hardware}}
    & 2 $\times$ 268 MHz  & \multirow{2}{*}{128MB}  & \multirow{2}{*}{1000MB} & ARM11MPCore & (ARM11) \\
    handheld video game console
    & 134 MHz  &                         &                         & ARM946      & (ARM9) \\
    \hline
    Raspberry Pi Zero
    \tablefootnote{\url{https://www.raspberrypi.org/products/pi-zero/}}
    & 1100 MHz & 512 MB  & External  & BCM2835 & (ARM11) \\
    \hline
    Samsung Galaxy S4
    \tablefootnote{\url{http://www.gsmarena.com/samsung_i9500_galaxy_s4-5125.php}}
    & 4 $\times$ 1600 MHz
    & \multirow{2}{*}{2 GB}
    & \multirow{2}{*}{ 16 / 32 GB}
    & \multirow{2}{*}{Exynos 5410 Octa}
    & (ARM Cortex-A15) \\
    smartphone & 4 $\times$ 1200 MHz & & & & (ARM Cortex-A7) \\
    \hline
    \end{tabular}
\end{center}
\end{table*}
\fi

All 32-bit ARM processors feature 16 general-purpose registers, denoted
\texttt{r0-r15}. Registers \texttt{r13-r15} have special names and usage
models. \ifnotabridged Table~\ref{tbl:arm-registers} lists each register and its corresponding
usage model. \fi These registers, including the program counter (\texttt{pc}) can be
accessed directly. Cortex-M processors implement two stacks, the \emph{Main}
stack and \emph{Process} stack. The stack pointer (\texttt{sp})
is \emph{banked} between processor modes, i.e., multiple copies of a register
exists in distinct register banks. Not all registers can be seen at once; the
register bank in use is determined by the current processor mode. Register banking
allows for rapid context switches when dealing with processor exceptions and
privileged operations. Application software on Cortex-M processor executes in
\emph{Thread} mode where the current stack is determined by the
\emph{stack-pointer select} (\texttt{spsel}) register. When the processor
executes an exception it enters the \emph{Handler} mode. In Handler mode the
processors always uses the Main stack. 
\ifabridged
When executing in Handler mode, the \emph{Interrupt Program Status Register}
(\texttt{ipsr}) holds the exception number of the exception being handled. The
\texttt{ipsr} may only be read using a \texttt{mrs} instruction used to access
ARM system register, and is only updated by the processor itself on exception
entry and exit (see \emph{Exception behaviour} in
Section~\ref{sec:implementation}).
\else

The \emph{Combined Program Status Register} \texttt{xpsr} is a 32-bit special
purpose register that is a combination of the logical \emph{Application}
(\texttt{apsr}), \emph{Execution} (\texttt{epsr}), and \emph{Interrupt}
(\texttt{ipsr}) Program Status Registers.  The \texttt{apsr} contains status
bits accessible to application-level software. The \texttt{epsr} contains the
execution state bit which describes the instruction set, which on Cortex-M CPUs
is only set to Thumb mode. When executing in Handler mode, the \texttt{ipsr}
holds the exception number of the exception being handled. The \texttt{ipsr} may
only be read using a \texttt{mrs} instruction used to access ARM system
registers, and is only updated by the processor itself on exception entry and
exit (see \emph{Exception behaviour} in Section~\ref{sec:implementation}).
Table~\ref{tbl:arm-special-registers} provides a non-exhaustive list of special
registers and their usage.
\fi

\ifnotabridged
\begin{table}[!tb]
  \begin{center}
    \caption{ARM General Purpose Registers~\cite{ARM-AAPCS}}
    \label{tbl:arm-registers}
    \begin{tabular}{ l l }
      \hline
      \multicolumn{1}{c}{\textbf{Register}} &
      \multicolumn{1}{c}{\textbf{Usage model}}                        \\
      \hline
      \texttt{r0 - r3}   &  Argument / scratch register               \\
      \texttt{r4 - r8}   &  Variable register (callee saved)          \\
      \texttt{r9}        &  Platform register                         \\
      \texttt{r10 - r11} &  Variable register (callee saved)          \\
      \texttt{r12 (ip)}  &  Intra-Procedure-call scratch register     \\
      \texttt{r13 (sp)}  &  Stack Pointer                             \\
      \texttt{r14 (lr)}  &  Link Register (subroutine return address) \\
      \texttt{r15 (pc)}  &  Program Counter                           \\
      \hline
    \end{tabular}
  \end{center}
\end{table}

\begin{table}[!tb]
  \begin{center}
    \caption{ARM Special Purpose Registers~\cite{ARMv8-M}}
    \label{tbl:arm-special-registers}
    \begin{tabular}{ l l }
      \hline
      \multicolumn{1}{c}{\textbf{Register}} &
      \multicolumn{1}{c}{\textbf{Usage model}}                        \\
      \hline
      \texttt{spsel}   &  Stack-Pointer Select Register               \\
      \texttt{xpsr}    &  Combined Program Status Register            \\
      \texttt{apsr}    &  Application Program Status Register         \\
      \texttt{epsr}    &  Execution Program Status Register           \\
      \texttt{ipsr}    &  Interrupt Program Status Register           \\
      \hline
    \end{tabular}
  \end{center}
\end{table}
\fi

\ifnotabridged
\paragraph{The Thumb instruction set.} ARM Cortex-M series processors utilize the
\emph{Thumb} instruction set, which is a subset of common commands found in the
32-bit RISC instruction set used in more powerful ARM Cortex-A series of
application processors. Thumb is a fixed-length 16-bit instruction set, where
each instruction is a compact shorthand for an instruction found among 32-bit
ARM instructions. Encoding a program in thumb code is around 25\% more
size-efficient than its corresponding 32-bit encoding. Modern ARM processors
extend the Thumb instruction set with 32-bit instructions, e.g. to achieve a
larger range or relative branch destinations. These new instructions are
distinct from those used in the 32-bit ARM instruction set, and may be
intermixed with 16-bit Thumb instructions.  The resulting variable-length
instruction set is referred to as \emph{Thumb-2}\footnotemark. Unlike the 32-bit
ARM instructions, which are encoded as 32-bit words, 32-bit Thumb-2 instructions
are encoded as two consecutive 16-bit half-words. The improved code density
compared to the 32-bit ARM instruction set makes Thumb better suited for
embedded systems where code footprint is often an important consideration due to
restricted memory bandwidth and memory cost. On ARM application cores, both
32-bit ARM and variable length Thumb instruction sets are supported and
interwork freely. Cortex-M processors only support Thumb and Thumb-2
instructions. Attempts to change the instruction execution state to 32 bit mode
causes a processor exception.

\footnotetext{For the remainder of this article, we use the terms \emph{Thumb}
and \emph{Thumb-2} interchangeably when referring to machine code that may
contain both 16-bit Thumb and 32-bit Thumb-2 instructions.}
\fi

\paragraph{ARM calling standard.} As with all processors, ARM provides a
\emph{calling standard} that compiler manufacturers should use to resolve
subroutine calls and returns in an interchangeable manner. In
programs conforming to the ARM Architecture Procedure Call Standard
(AAPCS)~\cite{ARM-AAPCS} subroutine calls may be performed either through a
\textbf{B}ranch with \textbf{L}ink (\texttt{bl}) or \textbf{B}ranch with
\textbf{L}ink and e\textbf{X}change (\texttt{blx}) instruction. These
instructions load the address of the subroutine to the \texttt{pc} and the
return address to the \emph{link register} (\texttt{lr}). ARM processors do not
provide a dedicated return instruction. Instead, a subroutine return is
performed by writing the return address to the program counter \texttt{pc}.
Hence, any instruction that can write to the \texttt{pc} can be leveraged as an
effective return instruction. Two common effective return instructions are
\texttt{bx lr} and \texttt{pop \{\ldots, pc\}}. The \texttt{bx lr} instruction
performs a branch to the return address stored in the link register \texttt{lr}.
The \texttt{pop \{\ldots, pc\}} in a subroutine epilogue loads the return
address from the stack to the \texttt{pc}. The former is typically used in
\emph{leaf routines}, which do not execute procedure calls to other routines.
The latter is typically preceded by a \texttt{push \{\ldots, lr\}} instruction
in the subroutine prologue, which in a \emph{non-leaf routine} stores the return
address in \texttt{lr} (possibly along with other registers that need to be
saved) on the stack in preparation for calls to other routines.

\subsection{TrustZone-M} \label{sec:tzm}
\noindent

\emph{TrustZone-M}~\cite{Yiu15,ARMv8-M} (TZ-M) is a new hardware security technology
present in the ARMv8-M architecture. In terms of functionality, it replicates the
properties of processor supported isolation and priority execution provided by
TrustZone-enabled Cortex-A application processors (TZ-A), but their respective
architectural realizations differ significantly. Both TZ architectures
expose a set of \emph{secure state} non-privileged and privileged processor
contexts beside their traditional \emph{non-secure state}
counterparts\footnotemark. In both TZ variants the memory management is
extended to enable splitting the device's physical memory into secure and
non-secure regions.
\ifnotabridged Fig.~\ref{fig:tzm-states} shows the relationship between
the traditional Thread and Handler modes (on the left hand side) and their new
secure state counterparts.
\fi

\footnotetext{Also referred to as the \emph{secure world} and \emph{normal
world}.}

\ifnotabridged
\begin{figure}[tb] \centering
\includegraphics[width=.8\hsize]{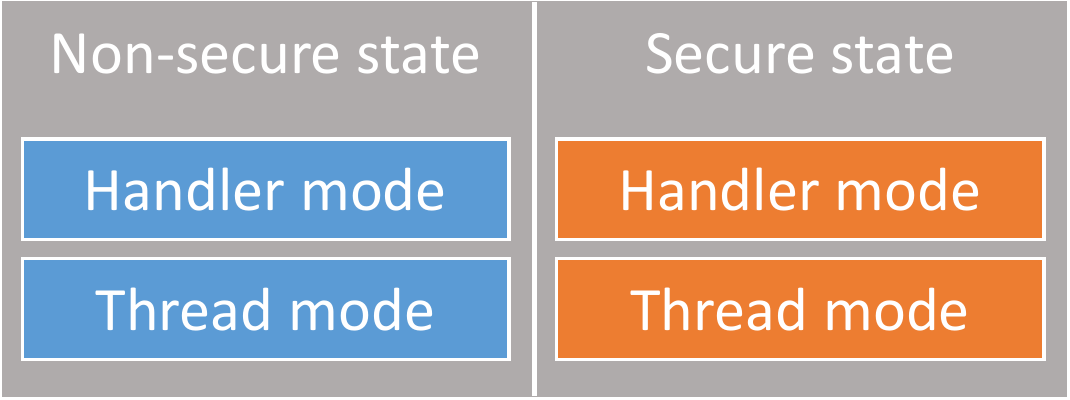} \caption{ARMv8-M CPU
states~\cite{Yiu15}} \label{fig:tzm-states} \end{figure}
\fi

In TZ-M, the only general purpose registers banked between the non-secure and
secure states are the \texttt{sp} registers used to address the Main and Process
stacks. \ifnotabridged Fig.~\ref{fig:tzm-registers} illustrates the set of registers in a
ARMv8-M equipped with TZ-M. The \texttt{MSP\_ns} and \texttt{PSP\_ns} represent
the Main and Process stack pointer in non-secure state, whereas the
\texttt{MSP\_s} and \texttt{PSP\_s} represent the corresponding stack pointers
in secure state. \fi The remaining general purpose registers are shared (not banked)
between the non-secure and secure states. In practice this means that the secure
state software is responsible for sanitizing any sensitive information held in
any general purpose registers during a transition from secure to non-secure
state\ifnotabridged\footnotemark.

\footnotetext{In addition any registers denoted callee save in the
AAPCS~\cite{ARM-AAPCS} must be stored and restored by the secure state software
upon secure state entry and exit respectively.}
\fi

\ifnotabridged
\begin{figure}[tb]
  \centering
    \includegraphics[width=.8\hsize]{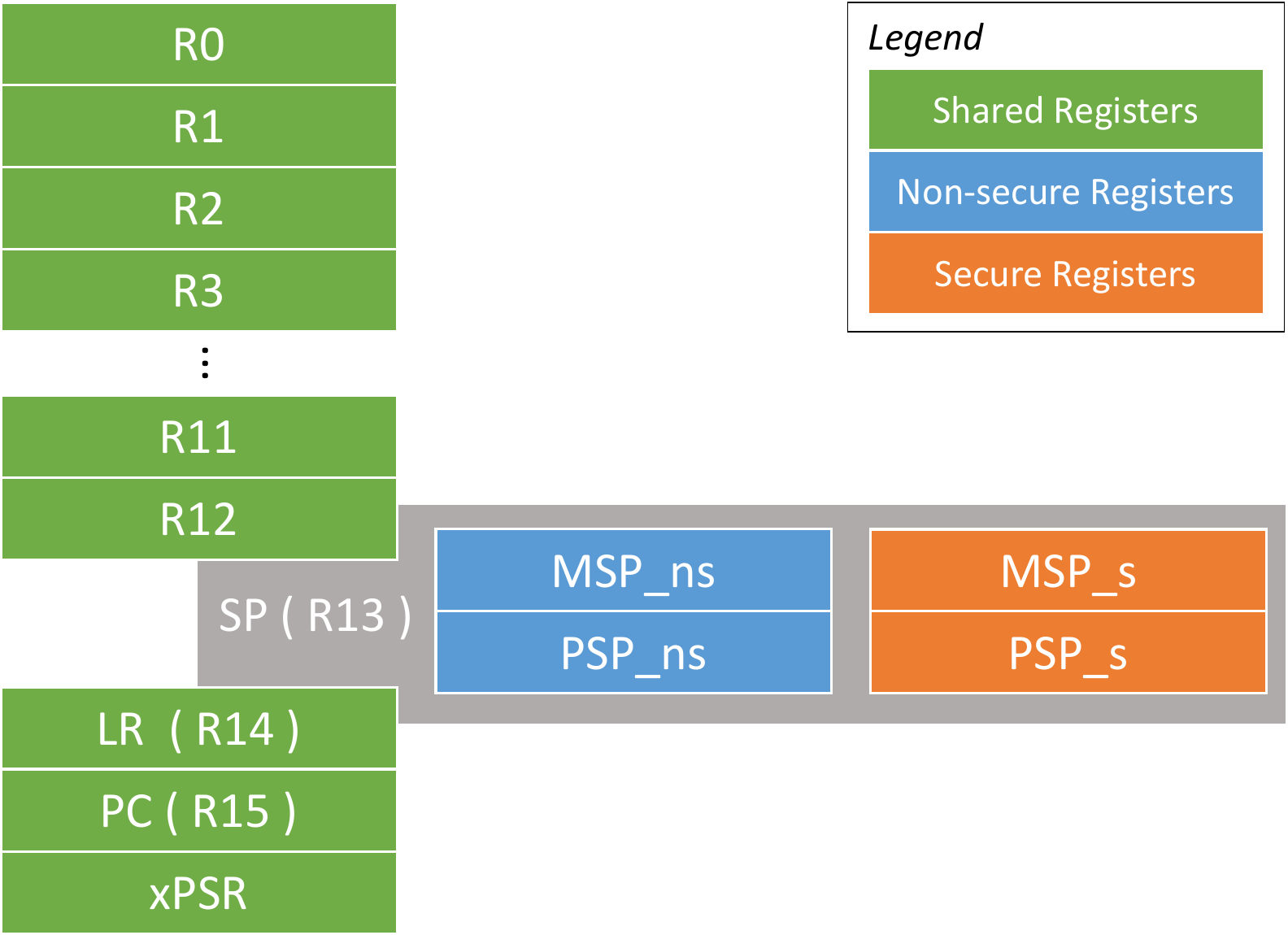}
  \caption{ARMv8-M registers~\cite{Yiu15}}
  \label{fig:tzm-registers}
\end{figure}
\fi

\begin{figure*}[htbp]
  \centering
  \ifabridged 
    \includegraphics[width=.9\hsize]{figures/tzm_secure_call_bw.pdf}
  \else
    \includegraphics[width=\hsize]{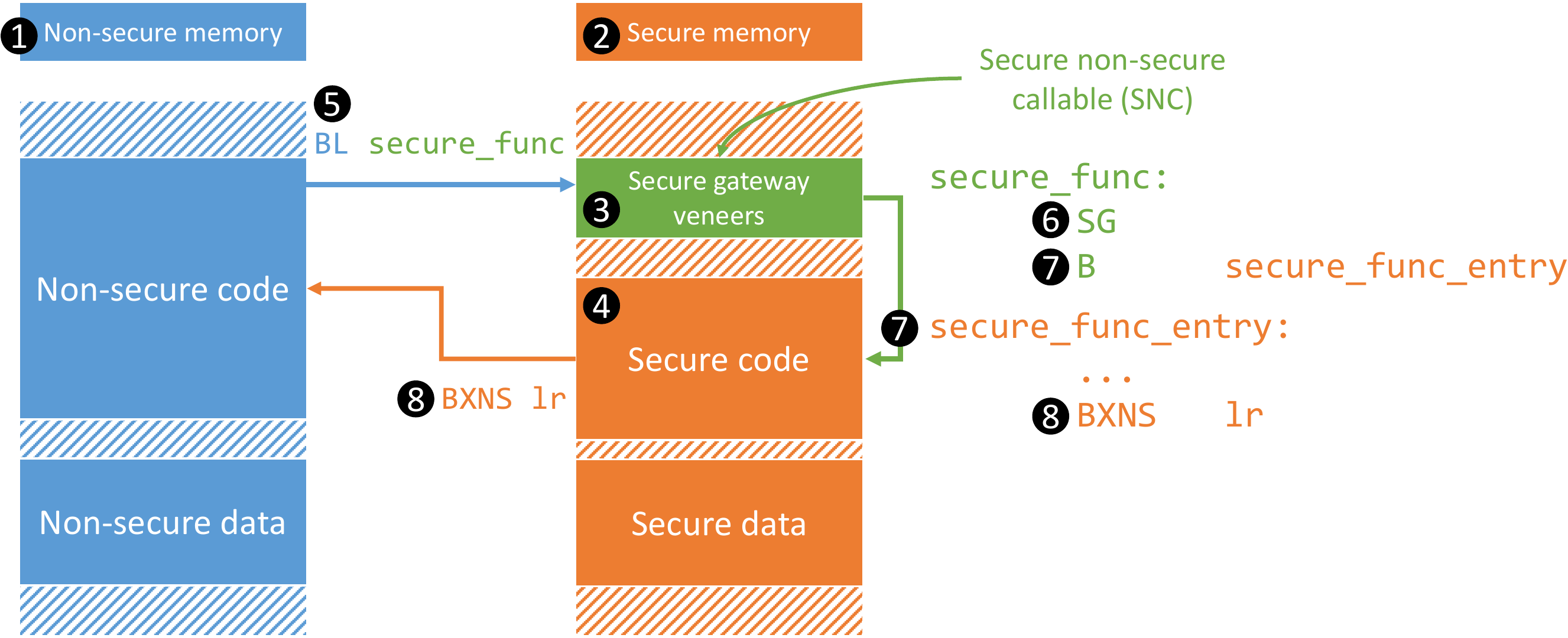}
  \fi
  \caption{ARMv8-M secure state call~\cite{Yiu15}}
  \label{fig:tzm-secure-call}
\end{figure*}

In TZ-A the entry to the secure state occurs via a dedicated hardware exception
and the context switch is performed by the exception handler known as the Secure
Monitor. In TZ-M the division between the secure and non-secure states is
instead based on a memory map set up during device initialization which
assigns specific regions of memory as either secure or non-secure. The
transitions between secure and non-secure state occur \emph{automatically} as
the flow of execution is transferred from program code in non-secure memory to
secure memory (and vice versa). Where in TZ-A the entry into the secure
state typically has to manage VM and MMU configuration at the expense of
thousands of processor cycles, TZ-M is geared towards embedded processors with
no virtual memory support (at most a MPU). In TZ-M a switch of security state
only takes a few processor cycles including a pipeline flush~\cite{Yiu15}.

The hardware support for the division of memory into secure and non-secure regions
in ARMv8-M is a Secure Attribution Unit (SAU) inside the processor. The SAU is
configurable while the processor executes in secure state. External interrupts
may be routed to either non-secure state exception handlers, or secure state
exception handlers based on the SAU configuration.
Fig.~\ref{fig:tzm-secure-call} denotes a typical memory layout for a TZ-M
equipped device. Each memory region known to the SAU may be declared as either
\emph{Non-Secure} (\dOne), \emph{Secure} (\dTwo) or \emph{Secure Non-Secure
Callable} (NSC \dThree). While Secure memory contains the secure program image
and data, the NSC memory contains \emph{secure gateway veneers}\footnotemark,
i.e., branch instructions (\dSeven) which point to the actual subroutine code in
Secure memory (\dFour). The purpose of the NSC is to prevent non-secure program
code to branch into invalid entry points in secure program code (such as into the
middle of a function, as is often done in atleast ROP). To this end, the ARMv8-M
instruction set also introduces a \emph{Secure Gateway} (\texttt{sg}) instruction,
which is included in the beginning of each veneer (\dSix) and acts as a call
gate to the secure program code. From the non-secure program code a call to a
secure subroutine is performed using a regular \texttt{bl} instruction (\dFive),
targeting the corresponding veneer in the NSC. Calls targeting a memory address
in the NSC will automatically cause a context switch to secure state, and the
processor will validate that the call targets a valid entry point with a
\texttt{sg} instruction. In particular, calls from non-secure state calling
secure memory outside the NSC, or non-\texttt{sg} instructions in the NSC will
fail in a \emph{Secure Fault}, a new type of hardware exception which always
traps into secure state. Secure subroutines return by executing a \texttt{bxns lr}
instruction (\dEight), which otherwise behaves like a return through \texttt{bx
lr}, but additionally switches the processor to non-secure state.

\footnotetext{\url{http://www.keil.com/support/man/docs/armclang_link/armclang_link_pge1444644885613.htm}}



\section{Problem Statement} \label{sec:cfi}

\subsection{Code-Reuse Attacks on ARM} \label{sec:rop}
\noindent

Code-reuse attacks are a class of software exploits that allow attackers to
execute arbitrary code on a compromised device, even in the presence of hardware
countermeasures against code injection, such as W$\oplus$X~\cite{HP-DEP}. In a \emph{return-to-libc} attack~\cite{Peslyak97}, the subroutine
return address on the call stack is replaced by the address of an entry point to
a subroutine in the executable memory of a process. 
The technique has been generalized into \emph{Return-Oriented Programming}~\cite{Shacham07} (ROP) for
the x86 architecture, which has since become the exploitation technique of
choice for modern memory-safety vulnerability attacks. Subsequently ROP has been
extended to various other CPU architectures~\cite{Buchanan08,Francillon08,Checkoway09},
including ARM microprocessors~\cite{Kornau09}.

\ifnotabridged
In a ROP attack, the attacker arranges the call stack to point to short
sequences of instructions in the executable memory of the victim program. The
instruction sequences, commonly referred to as \emph{gadgets}, are chosen so
that each ends in a return instruction, (i.e., a \texttt{pop \{\ldots, pc\}} or
\texttt{bx lr} in the ARM architecture) which, as long as the attacker has
control of the return address, causes each gadget to be executed in sequence. The
crucial advancement of ROP compared to return-to-libc attacks is that, given a
suitable set of gadgets, \emph{Turing complete} code execution \emph{without}
code injection can be achieved. Hence, a standard result in ROP work is to
show the presence of a Turing complete set of gadgets in the victim program
executable memory~\cite{Shacham07,Buchanan08,Francillon08,Checkoway09,Kornau09}.
Usually such a set of gadgets is sought from a commonly used shared library,
such as libc or standard Windows DLLs, to demonstrate the applicability of the
attack in arbitrary programs that link to this standard library.  However, it
should also be noted that a great deal of work prior to 2007 shows that even
without a Turing complete gadget set, it is possible to leverage control of the
stack to manipulate program execution in meaningful
ways~\cite{Wojtczuk98,Nergal01,Krahmer05}.
\fi

Many code-reuse attacks on x86 platforms use unintended instruction sequences
found by performing a branch into the middle of otherwise benign instructions.
Such unintended sequences cannot be formed in the 32-bit ARM, or in the 16-bit
Thumb instruction sets where branch target alignment is enforced on instruction
load, and hence may only target the intended instruction stream. However, the
presence of both 32-bit and 16-bit instructions in Thumb-2 code introduces
ambiguity when decoding program code from memory. When decoding Thumb-2
instructions, ARM processors still enforce 2-byte alignment on instruction
fetches, but the variable-length encoding allows the second half-word in a
32-bit Thumb-2 instruction to be interpreted as the first half-word of an
unintended instruction. Such unintended instructions have been successfully
utilized in prior work~\cite{Le11,Lian15} to exploit ARM code.

\ifnotabridged
Various proposed defenses against ROP have attempted to leverage properties of
the gadgets executed during a ROP attack. The attacks alter the program flow in
at least two ways: 1) they contain many return instructions, occurring only a few
instructions apart, and 2) the returns unwind the call stack without a
corresponding subroutine call.
\fi

It has been shown that, on both x86 and ARM, it is also possible to perform ROP
attacks \emph{without the use of return instructions}~\cite{Checkoway10} in what
has become to be known as \emph{Jump-Oriented Programming} (JOP). On ARM
platforms, JOP can be instantiated using indirect subroutine calls.

\subsection{Control-Flow Integrity} \label{sec:cfi}
\noindent

A well known approach to address code-reuse attacks is enforcing the
\emph{Control-Flow Integrity} (CFI) of the code. The execution of any program
can be abstractly represented as a \emph{Control-Flow Graph} (CFG), where nodes
represent blocks of sequental instructions (without intervening branches), and
edges represent control-flow changes between such nodes (branch instructions).
CFI enforcement strives to ensure that the execution of the programs conforms to
a legitimate path in the program's CFG. CFI builds on the assumption that
program code in memory is not writable (i.e., that memory pages can be marked
W$\oplus$X) as a countermeasure against code injection attacks. Code
immutability allows CFI checks to be omitted for nodes in the CFG that end in
direct branch instructions~\cite{Abadi09,Erlingsson06}, i.e., branches with a
statically determined target offset. As a result, CFI is typically applied to
nodes in the CFG that end in an indirect branch. Indirect branches are typically
emitted for switch-case statements, subroutine returns, and indirect calls
(subroutine calls to dynamic libraries, calls through function pointers, e.g.
callbacks, as well as C++ virtual functions).

While the construction of the CFG can occur through static inspection of the
program binary, the actual enforcement of CFI must occur at
runtime. In inlined CFI enforcement the checks that validate control-flow changes are interspersed with the original program code at subroutine call sites, as
well as in the subroutine prologue and epilogue\ifnotabridged\footnotemark\fi.
The insertion of these checks can be achieved through compiler
extensions~\cite{Chiueh01}, or by binary machine-code rewriting. Binary
instrumentation that adds additional instructions to a pre-built program binary
by necessity modifies the memory layout of the code, and hence will require
memory addresses referenced by the program to be adjusted accordingly.

\ifnotabridged
\footnotetext{The subroutine prologue is a compiler generated sequence of
instructions in the beginning of each subroutine which prepares the stack and
registers for use within the subroutine. Similarly, the subroutine epilogue
appears at the end of the subroutine, which restores the stack and registers
to the state prior to the subroutine call.}
\fi

Traditional ROP targets return instructions that read the return address
off the program stack. A well known technique to enforce that subroutine
returns target the original call site is the notion of a \emph{shadow call
stack}~\ifabridged{\cite{Dang15}\else\cite{Abadi09,Chiueh01,Frantzen01,Giffin02,Giffin04,Erlingsson06,Nebenzahl06,Prasad03,Francillon09,Davi11,Carlini15,Intel-CET}\fi. The shadow
call stack is used to hold a copy of the return address. On subroutine
return the return address on the shadow call stack is compared to the
return address on the program stack. If they match, the return proceeds as
usual. A mismatch in return addresses on the other hand indicates a failure of
CFI and triggers an error which terminates the program prematurely. Recent
results show that, in fact, shadow stacks are essential for the security of
CFI~\cite{Carlini15}.

\subsection{CFI Challenges for Microcontrollers} \label{sec:cfi}
\noindent

We identify the following challenges in realizing CFI protection
on IoT devices:
\begin{itemize}
 \itemsep0em
\item \textbf{Interrupt awareness}: Since the software to be
  protected is a single, interrupt-driven bare-metal program, the CFI
  scheme needs to handle both interruptible code, as well as
  execution in interrupt contexts. To the best of our knowledge, no
  existing CFI scheme meets this requirement.
\item \textbf{Hardware-based shadow stack protection}: Protection of
  shadow stack must leverage lightweight hardware-based trust anchors
  like TrustZone-M. The code size and performance overhead of purely
  software-based CFI is prohibitive on resource constrained devices
  and techniques for general purpose computing devices often rely on
  hardware (such as x86 segmentation support~\cite{Abadi09}) that is
  unavailable in simple MCUs.
\item \textbf{Layout-preserving instrumentation}: Since software for
  MCUs is commonly deployed as monolithic firmware images with strict
  size requirements, CFI instrumentation must preserve memory layout
  of the image so as to avoid extensive rewriting and to minimize the
  increase in code size.
\item \textbf{On-device instrumentation}: To avoid having to rely on
  the developer (or some other external entity) to perform the required
  instrumentation, the CFI scheme must be amenable to on-device
  instrumentation. 
\end{itemize} 

\subsection{Adversarial Model} \label{sec:adversary-model}
\noindent

We consider a powerful adversary with arbitrary read-access to code memory and arbitrary read-write access to data memory of the non-secure state program. This model accounts for buffer overflows or other memory-related vulnerabilities (e.g. an externally controlled format string\footnotemark) that, in practice, would allow adversaries to gain such capabilities. The adversary cannot modify code memory, a property that is achievable even on MCU class systems through widespread countermeasure against code injection (e.g. MPU-based W$\oplus$X). Nevertheless, arbitrary read-access necessitates a solution that is able to withstand information disclosure (the strongest attack scenario in Dang et al.’s~\cite{Dang15} evaluation of prior work on CFI). Our threat model is therefore similar to previous work on CFI, but we also consider an even stronger adversary who can exploit interrupt handling to undermine CFI protection. 

This model applies even when an attacker is in active control of a module or thread within the same address space as the non-secure state program, such as gaining control of an unprotected co-processor on the \emph{System-On-Chip} (SoC). However, the adversary lacks the ability to read or modify memory allocated to the secure state software. 

\footnotetext{CWE-134: Use of Externally-Controlled Format String\\
\url{https://cwe.mitre.org/data/definitions/134.html}}

In this work, we do not consider \emph{non-control data} attacks~\cite{Szekeres13}
such as \emph{Data-Oriented Programming}~\cite{Hu16}. This class of
attacks can achieve privilege escalation, leak security sensitive data or even
Turing-complete computation by corrupting memory variables that are not directly
used in control-flow transfer instructions. This limitation also applies to
prior work on CFI.

\section{\tzmcfisectitle}
\label{sec:tzmcfi}

We now present CaRE (Call and Return Enforcement), our solution for ensuring control-flow integrity.
\tzmcfi specifically targets constrained IoT devices, which are expected to stay active in the field for a prolonged time and operate unattended with network (Internet) connectivity, possibly via IoT gateways. This kind of deployment necessitates the incorporation of software update mechanisms to fix vulnerabilities, update configuration settings and add new functionality.

We limit our scope to small, more or less bare-metal IoT devices. The system software is deployed as monolithic, statically linked firmware images. The secure and non-secure state program images are distinct from each other~\cite{ARM-LLVM}, with the secure state software stack structured as a library. The configuration of the SAU and the secure state program image is performed before the non-secure code is started. The entry to the secure state library happens through a well-defined interface describing the call gates available to non-secure software. Functions in the secure state are synchronous and run to completion unless interrupted by an exception. The system is interrupt-driven, reacting to external triggers. While it is possible that the non-secure state software is scheduled by a simple Real-Time Operating System (RTOS), the secure state software does not have separate scheduling or isolation between distinct software components for the simple reason that the device is single-purpose rather than a platform for running many programs from many stakeholders in parallel. Even when an RTOS is present, it is seldom necessary for non-secure state code to support dynamic loading of additional code sequences.

\subsection{Requirements}
\label{sec:requirements}

Given the above target deployment scenario, we formulate the following requirements that \tzmcfi should meet:

\begin{requirement} \label{req:cfi}
  It must reliably prevent attacks from redirecting the flow of execution of the non-secure state program.
\end{requirement}
  
\begin{requirement} \label{req:c-code}
  It must be able to protect system software written in standard C and assembler conformant to the AAPCS.
\end{requirement}

\begin{requirement} \label{req:code-footprint}
  It must have minimal impact on the code footprint of the non-secure state program.
\end{requirement}

\begin{requirement} \label{req:performance}
  Its performance overhead must be competitive compared to the overhead of software-based CFI schemes.
\end{requirement}

We make the following assumptions about the target device:

\begin{assumption} \label{as:trust-anchor}
  A trust anchor, such as TZ-M, which enables isolated code execution and secure storage of data at runtime is available.
\end{assumption}

\begin{assumption} \label{as:bootstrap}
  All (secure and non-secure) code is subject to a \emph{secure boot} sequence that prevents tampering of program and update images at rest. This bootstrap sequence itself is not vulnerable to code-reuse attacks, and routines in the bootstrap code are not invoked again after the device startup completes. 
\end{assumption}

\begin{assumption} \label{as:rom}
  All code is non-writable. It must not be possible for an attacker to modify the program code in memory at runtime.
\end{assumption}

\begin{assumption} \label{as:dep}
  All data is non-executable. It must not be possible for an attacker to execute data as it were code. Otherwise, an attacker will be able to mount code injection attacks against the device.
\end{assumption}

Assumption~\ref{as:trust-anchor} is true for commercial off-the-shelf ARMv8-M MCUs. There also exist several research architectures, such as SMART~\cite{Defrawy12}, SANCUS~\cite{Noorman13}, and Intel's TrustLite~\cite{Koeberl14} that provide equivalent features. Assumption~\ref{as:bootstrap} is true 
for currently announced ARMv8-M SoCs~\footnotemark. Assumptions~\ref{as:rom} and \ref{as:dep} are in line with previous work on CFI and can be easily achieved on embedded devices that are equipped with MPUs. These assumptions can be problematic in the presence of self-modifying code, runtime code generation, and unanticipated dynamic loading of code. Fortunately, most embedded system software in MCUs is typically statically linked and written in languages that compile directly to native code. Even when an RTOS is present, it is seldom necessary for non-secure state code to support dynamic loading of additional code sequences.

\footnotetext{\url{https://www.embedded-world.de/en/ausstellerprodukte/embwld17/product-9863796/numicro-m2351-series-microcontroller}}

\subsection{Architecture}
\label{sec:architecture}

\begin{figure}[!tb]
  \centering
  \ifabridged  
    \includegraphics[width=\hsize]{figures/tzmcfi_overview_bw}
  \else
    \includegraphics[width=\hsize]{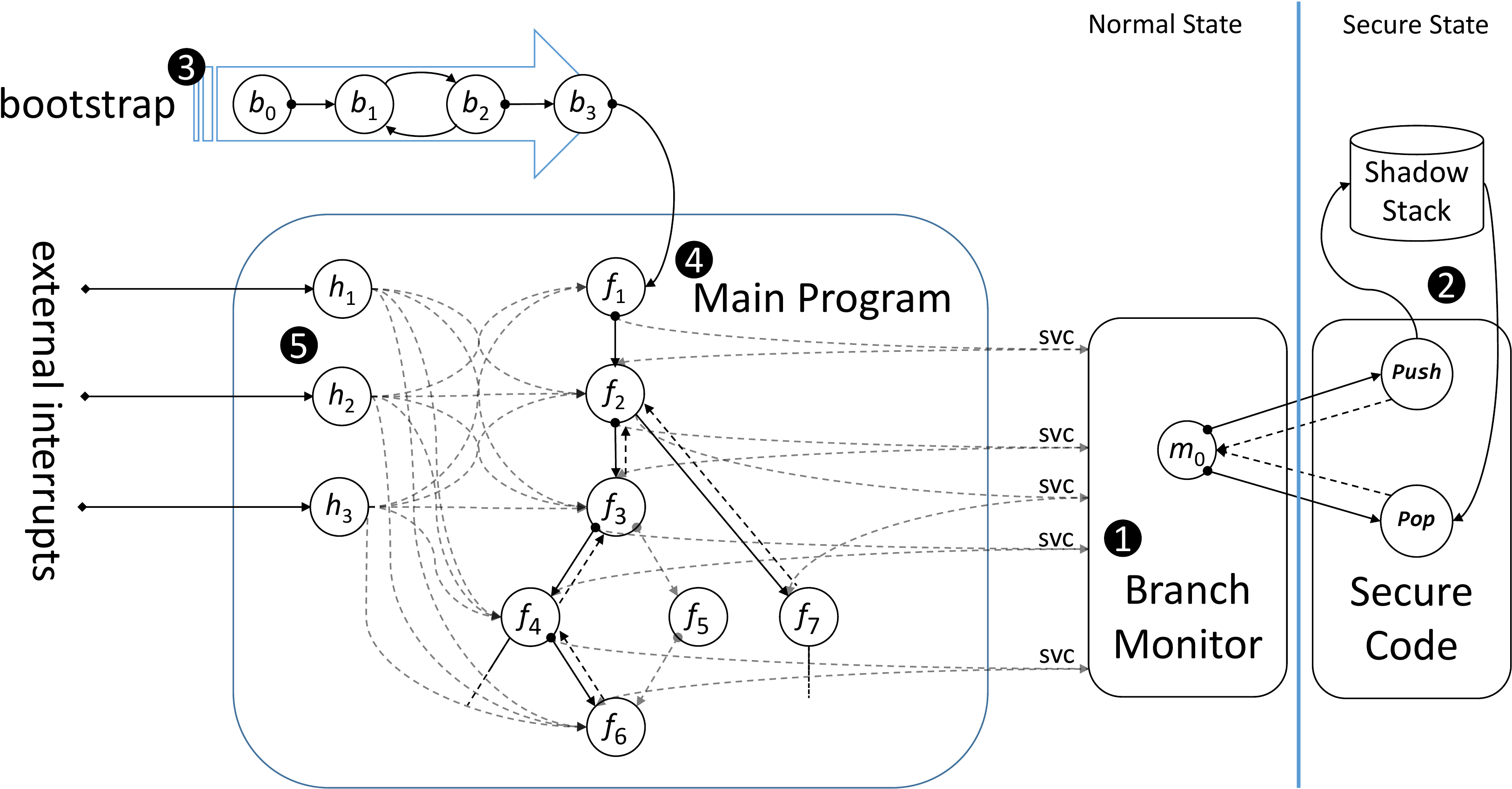}
  \fi
  \caption{\tzmcfi overview}
  \label{fig:tzmcfi-architecture}
\end{figure}

Our design incorporates protection of a shadow call stack on low-end ARM embedded devices featuring TZ-M. The shadow call stack resides in secure memory, and is only accessible when the processor executes in the secure state. We also propose a {\it layout-preserving} binary instrumentation approach for Thumb code, with small impact to code footprint, and an opportunity for {\it on-device instrumentation} as part of code installation. The main aspect of this property is that the binary program image is rewritten without affecting its memory layout. Fig.~\ref{fig:tzmcfi-architecture} shows an overview of the \tzmcfi architecture.

The premise for \tzmcfi is instrumentation of non-secure state code in a manner which removes all function calls and indirect branches and replaces them with dispatch instructions that trap control flow to a piece of monitor code, the \emph{\monitor} (\dOne), which runs in non-secure state. As a result, each subroutine call and return is now routed through the \monitor. The \monitor maintains the shadow stack by invoking secure functions (\dTwo) only callable from the \monitor, before transferring control to the original branch target. Other indirect branches, such as ones used to branch into switch case jump tables can be restricted by the \monitor to a suitable range and to target direct branches in jump table entries. Thus, the \monitor provides \emph{complete mediation} of instrumented non-secure state code.

Apart from the \monitor, the program image also contains bootstrap  routines (labeled $b_{n}$) that are used to initialize the runtime environment (\dThree). Such routines may initially need to operate without a stack and other memory structures in place, and as such are typically hand written in assembler. Due to these constraints, the bootstrap routines are likely to deviate from usual AAPCS conventions. In particular, all calls are not guaranteed to result in a subsequent matching return as fragments of bootstrap routines may simply be chained together until eventually transferring control to the named C entry point marking the beginning of main program code. On the other hand, the initialization code is typically not entered again after control has been transfered to the main function until the device is reset.

Hence, from the perspective of maintaining control-flow integrity, both the \monitor and bootstrap code exist outside benign execution paths encountered in the program during normal operation. Henceforth, we will refer to the code reachable from the main function as the \emph{main program}. The CFG nodes labeled $f_{n}$ in Fig.~\ref{fig:tzmcfi-architecture} represent the instrumented main program (\dFour). The main program and bootstrap code do not share any routines (Assumption~\ref{as:bootstrap}), even though routines belonging to one or the other may be interleaved in program memory. The main program code constitutes a \emph{strongly connected component} within the \emph{call graph}\footnotemark. This observation leads us to consider the main program code as a complete ensemble in terms of instrumentation target. It can include an RTOS and/or interrupt handlers. Interrupts handlers labeled $h_{n}$ (\dFive), with the exception of the supervisor call handler that hosts the \monitor, are considered to be part of the main program. Conceptually, interrupts may be reached from any node in the program's CFG. 

\footnotetext{A call graph is a control-flow graph which represents the calling relationships between subroutines in a program.}

By eliminating non-mediated calls and returns in the non-secure state main program, thus forcing each indirect branch through the \monitor, we can unequivocally eliminate control-flow attacks that utilize such branches.

\ifacmart  
\vspace{4\baselineskip}
\fi
\subsection{Instrumentation}
\label{sec:instrumentation}

The instrumentation must intercept all subroutine calls and returns. Furthermore, it should have minimal impact on code footprint. Prior shadow stack schemes either instrument the subroutine prologue and epilogue~\cite{Chiueh01,Dang15}, or the call site~\cite{Dang15}, pushing the return address to the shadow stack upon a subroutine call, and validating the return address on top of the shadow stack upon return. We propose an alternative approach which is layout-preserving.

In uninstrumented code, the target address of direct subroutine calls (i.e., \texttt{bl} instructions with immediate operands) are encoded as \texttt{pc}-relative offsets (i.e., signed integer values). In other words, the destination address depends on the location of the branch instruction in memory. During instrumentation, we calculate the absolute destination address, and store it in a data structure, called the \emph{branch table} which at runtime resides in read-only non-secure memory. Each destination address in this branch table is indexed by the memory address of the original branch instruction. The original branch instruction is overwritten with a \emph{dispatch instruction}, which, when executed, traps into the \monitor. At runtime, whenever an instruction rewritten in this fashion traps into the \monitor, the \monitor will lookup the destination address from the branch table, and redirect control flow to the original destination address.

In a similar manner, indirect branches corresponding to calls and effective returns are replaced with dispatch instructions. The destination address of the branches are only known at runtime, determined by a register value (\texttt{lr} in the case of effective returns), or by a return address stored on the program call stack, and hence do not influence the construction of the branch table during instrumentation.

To address JOP attacks, our CFI enforcement must also be able to determine legal call targets for indirect calls. In the case of indirect subroutine calls, the call target must be a valid entry point to the destination subroutine, i.e., the call must target the beginning of the subroutine prologue. The entry addresses are extracted from the symbol information emitted by the compiler for debug purposes. Further restriction of call targets is possible by means of static or dynamic analysis (see Section~\ref{sec:extensions}). Since \tzmcfi only requires the addresses of entry points, not the full debug information, the entry points are included in the software image in a \emph{call target table} on the device in a similar manner to the branch table. When an indirect call occurs, the \monitor will match the branch target against this record of valid subroutine entry points. 

In our implementation, we use the supervisor call \texttt{svc} instruction as the dispatch instruction, and place the \monitor in the supervisor call exception handler. The \texttt{svc} instruction has a number of desirable properties which make it suitable as a dispatch. Firstly, it allows for an 8-bit comment field, which is ignored by hardware, but can be interpreted in software, typically to determine the service requested. We exploit this comment field to identify the type of the original instruction, overwritten during the instrumentation (e.g. \texttt{bl}, \texttt{blx}, \texttt{pop \{pc\}} etc.). Secondly, the supervisor call handler executes at the highest exception priority, allowing us to pre-empt execution to the \monitor when instrumenting exception handlers. Lastly, because the \texttt{svc} in Thumb  instruction is a 16-bit instruction, it can be used for instrumenting both 32-bit and 16-bit instructions. When replacing 32-bit instructions, e.g., a Thumb-2 \texttt{bl} instruction with an immediate operand, we use the sequence \texttt{0xb000}, which corresponds to the opcode for \texttt{add sp, \#0} (effectively a NOP) as padding to fill the remaining 16 bits of the original \texttt{bl}.

\subsection{Implementation}
\label{sec:implementation}

\begin{figure*}[!tb]
  \centering
  \ifabridged  
    \includegraphics[width=\hsize]{figures/tzmcfi_instrumented_call_bw}
  \else
    \includegraphics[width=\hsize]{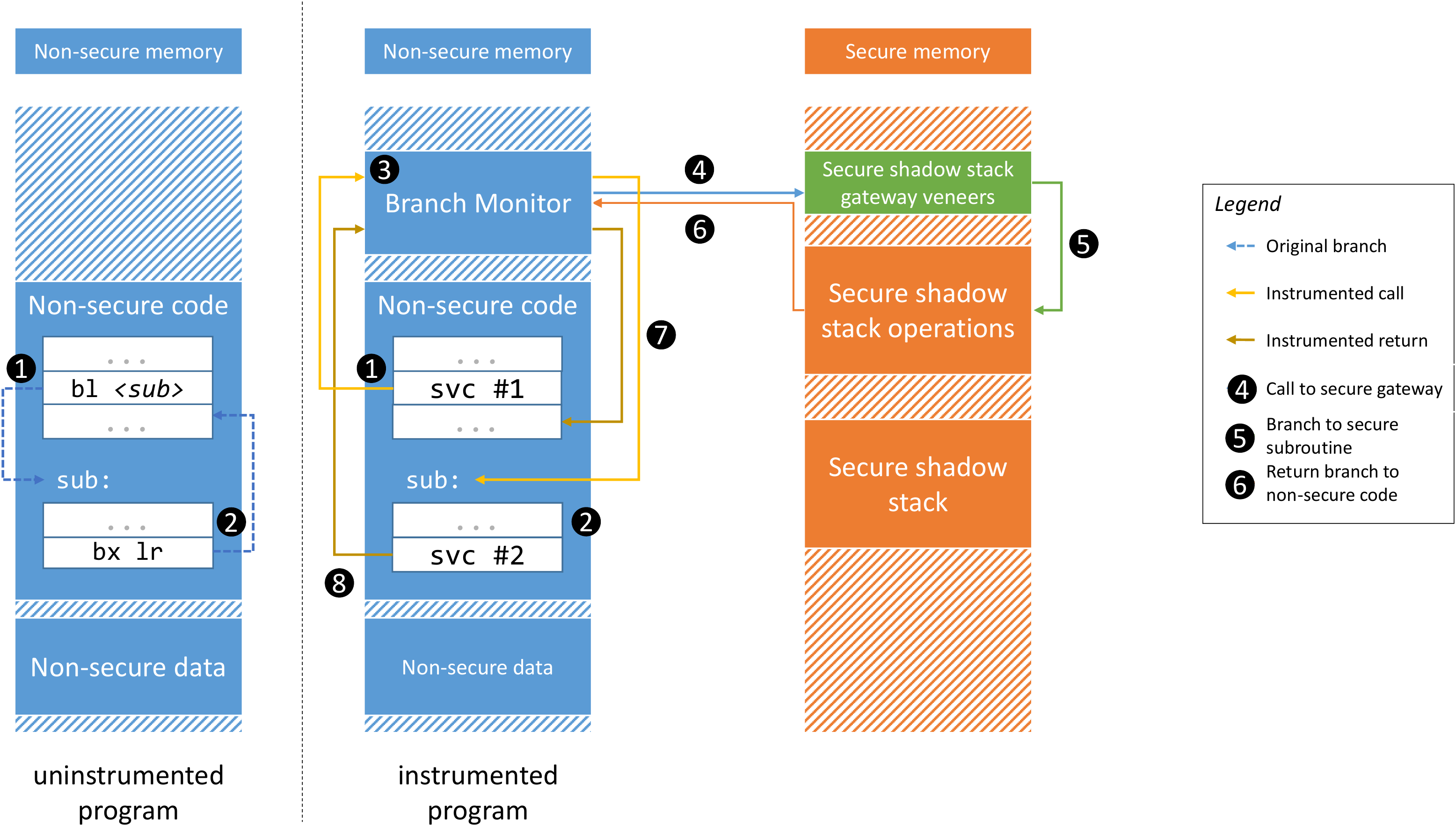}
  \fi
  \caption{\tzmcfi instrumented control flow}
  \label{fig:tzmcfi}
\end{figure*}

We implemented a proof-of-concept prototype of \tzmcfi on the ARM Versatile Express Cortex-M Prototyping System MPS2+ configured as a Cortex-M23 CPU.\ifnotabridged\footnotemark.

\footnotetext{We also tested our prototype on the CMSDK\_ARMv8MBL FastModel emulator included in version 1.3.0 of the Cortex Microcontroller Software Development Kit in Keil $\mu$Vision version 5.20.0.0. The prototype uses the  version 5.0.0-Beta4 of the Cortex Microcontroller Software Interface Standard (CMSIS). Binaries for instrumentation were produced using ArmClang v6.4 included in MDK-ARM version 5.20.}
\fi

We implemented a binary rewriter to perform the instrumentation on non-secure state binaries. It utilizes the Capstone disassembly engine\footnote{\url{http://www.capstone-engine.org/}} to identify control-flow instructions for rewriting.

Fig.~\ref{fig:tzmcfi} illustrates the altered control-flow changes. When a dispatch instruction is encountered in the program (\dOne), instead of taking a control-flow transfer directly to the original target (\dTwo), program execution is temporarily halted by a trap into the \monitor (\dThree). The \monitor will update the shadow stack maintained in secure memory by invoking secure shadow stack operations entry points in the gateway veneer (\dFour), which allow access to the secure state subroutines handling the actual update (\dFive). Upon completion, control is returned to the non-secure \monitor code (\dSix), which finally redirects control flow to the intended destination (\dSeven). The same sequence applies both for calls, and returns (\dEight).

\paragraph{\monitorsectitle.}

The \monitor is responsible for dispatching and validating control-flow transfers that occur during program execution. When invoked, it will first determine the reason for the trap based on the \texttt{svc} comment and trigger the corresponding branch handler routine within the \monitor. The routine updates the shadow stack accordingly (pushes return address on subroutine calls, pops and validates return address on subroutine returns) and redirects the control flow to the intended target. For branches corresponding to direct subroutine calls a branch table lookup is needed since the target of a call is not anymore evident from the dispatch instruction. For indirect calls, the \monitor verifies that each call targets a valid subroutine entry within the main program by looking up the target from the call target table.

As the \monitor executes in the supervisor call handler, the main stack contains a context state stack frame corresponding to the processor state at the point the supervisor call exception was taken (see Table~\ref{tbl:ex-stack-frame}). Control-flow redirection is triggered by manipulating stored \texttt{pc} and \texttt{lr} values in the context stack frame and performing an exception return from the \monitor (see below), which causes the processor to restore the context stack frame and resume execution from the address in the stored \texttt{pc}.

\begin{table}[!tb]
  \begin{center}
    \caption{Context state stack frame layout~\cite{ARMv8-M}}
    \label{tbl:ex-stack-frame}
    \begin{tabular}{ c | c | c }
      \multicolumn{1}{c}{\textbf{Offset}} &
      \multicolumn{1}{c}{\textbf{Stack Contents}} \\
      \cline{2-2}
      \texttt{0x1C} & \texttt{xpsr} & \multicolumn{1}{c}{} \\
      \cline{2-2}
      \texttt{0x18} & \texttt{pc}   & \\
      \cline{2-2}
      \texttt{0x14} & \texttt{lr}   & \\
      \cline{2-2}
      \texttt{0x10} & \texttt{r12}  & \\
      \cline{2-2}
      \texttt{0x0C} & \texttt{r3}   & \\
      \cline{2-2}
      \texttt{0x08} & \texttt{r2}   & \\
      \cline{2-2}
      \texttt{0x04} & \texttt{r1}   & \\
      \cline{2-2}
      \texttt{0x00} & \texttt{r0}   & $\leftarrow$ \texttt{sp} \\
      \cline{2-2}
    \end{tabular}
  \end{center}
\end{table}

\paragraph{Interrupt awareness.}

An important feature of M-class cores is their deterministic interrupt latency in part attributable to the fact that the context-switch, while entering the exception handler, is performed entirely in hardware. An instruction that triggers an exception, such as the \texttt{svc} used for supervisor calls, causes 1) the hardware to save the current execution context state onto a stack pointed to by one of the \texttt{sp} registers, 2) the \texttt{ipsr} to be updated with the number of the taken exception, and 3) the processor to switch into \emph{Handler mode} in which exceptions are taken. Table~\ref{tbl:ex-stack-frame} shows the layout of a typical stack frame created during exception entry\footnotemark. The value stored at offset \texttt{0x18} in the stack frame is the \texttt{pc} value at the point the exception was taken, and represents the return value from which program execution shall continue after the exception handler exits. To facilitate fast restoration of the saved context state, M-class processors support a special return sequence which restores the saved values on the stack into their corresponding registers. This sequence is known as an \emph{exception return} and occurs when the processor is in Handler mode, and a special \emph{Exception Return Value} (ERV) is loaded into the \texttt{pc} either via a \texttt{pop} instruction, or a \texttt{bx} with any register. ERVs are of the form \texttt{0xF}\textit{XXXXXXX}, and encode in their lower-order bits information about the current processor state and state before the current exception was taken. ERVs are not interpreted as memory addresses, but are intercepted by the processor as they are written to the \texttt{pc}. When this occurs, the processor will validate that there is an exception currently being handled, and that its number matches the exception number in the \texttt{ipsr}.  If the exception numbers match, the processor performs an exception return to the processor mode specified by the ERV, restoring the previous register state from the current stack, including the stored \texttt{pc}. This causes the processor to continue execution from the point in the program at which the exception was originally taken. When multiple exceptions are pending, lower priority exceptions may be \emph{tail-chained} which causes the processor to directly begin executing the next pending exception, without restoring the context state frame between exceptions.

\footnotetext{In Cortex-M processors that implement the floating point extensions, the context stack frame may also contain the values of floating point registers.}

Due to the fact that the context state stack frame contains a stored \texttt{pc} value that is restored on exception return, an exception handler with a vulnerability that allows an attacker to control the content of the context state frame on the stack constitutes a valid attack vector. This attack differs from a traditional ROP attack in that the attacker does not need to control the immediate \texttt{lr} value (which may reside only in the \texttt{lr} register), as during the execution of the exception handler \texttt{lr} contains merely the current ERV value. Instead, by manipulating the \texttt{pc} value in the context state stack frame, an attacker can cause an effective return from the exception handler to an arbitrary address. To avoid this, \tzmcfi needs to be interrupt aware, and accurately record the correct return address for the exception handler onto the shadow stack. However, exceptions (when enabled) may be triggered by events external to the main program execution, effectively pre-empting the main program at an arbitrary position in the code, even during the execution of another exception handler (assuming an exception of higher priority arriving concurrently).

To tackle this challenge this, we introduce \emph{exception trampolines}. When an exception is received, the trampoline determines the return address, stores it on the shadow stack, and then proceeds to execute the original exception handler. The exception trampolines can be instrumented in place by rewriting the non-secure state exception vector and replacing the address of each exception with the address of a corresponding exception trampoline, that ends in a fixed branch to the original exception handler. That address is the original exception vector entry.

Since \tzmcfi may interrupt the execution of another exception handler, we need to support a nested exception return, i.e. when the \texttt{pc} is being supplied with two consecutive return values in immediate succession. However, \texttt{pc} values in the \texttt{0xF0000000} - \texttt{0xFFFFFFFF} range are only recognized as ERVs when they are loaded to the \texttt{pc} either via a \texttt{pop} instruction, or a \texttt{bx} with any register (see Section~\ref{sec:thumb}). In particular, when an ERV is loaded to the \texttt{pc} as part of an exception return, it is instead interpreted as a memory address in an inaccessible range thus causing a hard fault in the processor. To overcome this, we also deploy \emph{return trampolines}, small fragments of instruction sequences that contain the different effective return instructions originally present in the program image prior to binary rewriting. When the \monitor returns from the supervisor call exception handler, it does so via the trampoline corresponding to the original return instruction. \ifnotabridged \ref{appx:sample-code} contains an excerpt from the \monitor, which illustrates the update of the \texttt{pc} and the \texttt{lr} through the stored context state frame and exception return via return trampolines. \fi

\section{Evaluation}
\label{sec:evaluation}

\subsection{Security Considerations}

\label{subsec:security}
A key consideration for the effectiveness of \tzmcfi is the ability of the \monitor to perform complete mediation of indirect control-flow events in untrusted non-secure state program code. After all, any branch instruction for which an adversary can control the destination may potentially be used to disrupt the normal operation of the program. In practice, it is not possible to completely eliminate all instructions that may act as indirect branches from the non-secure state program image. In particular, the bootstrap code, the \monitor itself and the return trampolines must remain uninstrumented. We argue that despite the \monitor and bootstrap code being uninstrumented, \tzmcfi is secure in terms of fulfilling Requirement~\ref{req:cfi}. We demonstrate this with the following reasoning.

\begin{claim}
In order to maintain the control-flow integrity of the non-secure state program it is \emph{sufficient} for the \monitor to mediate calls that occur within the strongly connected component of the main program's call graph.
\end{claim}

\noindent We base our reasoning on the following observations:

\begin{observation}
\label{obs:trusted-monitor}
  The secure state software stack, and the \monitor are trusted and cannot be disabled or modified.
\end{observation}

This follows from Assumptions~\ref{as:bootstrap} and \ref{as:rom}. A secure boot mechanism protects the program code at rest and read-only memory protects it from modification at runtime.

\begin{observation}
\label{obs:instr}
The main program has been instrumented in a manner which replaces \emph{all} subroutine calls and indirect branch instructions with \monitor calls.
\end{observation}

This follows simply from the operation of our instrumentation.

\noindent Based on these observations we formulate the following invariants:

\begin{invariant}
\label{inv:entry-points}
  Each subroutine within the main program has a fixed entry address that is the entry point for all control-transfer instructions (that are not returns) that branch to the subroutine.
\end{invariant}

\begin{invariant}
\label{inv:returns}
  All control-transfer instructions in the main program that act as effective returns target a previously executed call site within the main program.
\end{invariant}

Invariant~\ref{inv:entry-points} is true for all subroutines that are entered by control-transfer instructions where the destination address is an immediate operand that is encoded into the machine code instruction itself. This remains true after instrumentation as per Observation~\ref{obs:trusted-monitor} and \ref{obs:instr}, as the destinations addresses are replicated read-only in the branch table, and the control-flow transfer for instrumented calls is mediated by the \monitor. The entry address to an interrupt handler is the address recorded in the interrupt vector, and thus fixed, as interrupt handlers are not called directly from main program code.

As long as Invariant~\ref{inv:entry-points} holds control-flow transfers to an offset from the beginning of a subroutine are not possible. This includes branches that target 32-bit Thumb-2 instructions at a 16-bit offset\footnotemark, thus attempting to make use of the ambiguity in the Thumb-2 instruction set encoding.

\footnotetext{Half-word alignment for branch instruction target addresses is enforced by the hardware itself.}

Invariant~\ref{inv:returns} follows during benign execution from the structure of the program's call graph and Assumption~\ref{as:bootstrap}. It remains true after instrumentation, notably \emph{even} in the case the return addresses are compromised, because Observations~\ref{obs:trusted-monitor},~\ref{obs:instr} and Invariant~\ref{inv:entry-points} imply that the \monitor has complete mediation of control-flow transfers within the main program. Thus, the \monitor has the ability to enforce that no return may target an address from which a matching call site has not been observed.

Based on this, and given that no instrumented subroutine call will neither ever occur from the bootstrap code nor from the \monitor into the main program we may formulate the following corollaries:

\begin{corollary}
  No return within the main program may target the \monitor.
\end{corollary}

\begin{corollary}
  No return within the main program may target the initialization code.
\end{corollary}

Hence, as long as the \monitor can correctly mediate all immediate branches corresponding to subroutine calls and all indirect branch instructions within the main program, the call return matching performed by the \monitor enforces that no control-flow transfers to outside the main program occur as a result of mediated calls. \ifacmart\hfill\ensuremath{\square}\fi

We evaluated the effectiveness of our \monitor implementation by attempting to corrupt control-flow data on the stack through a buffer overflow introduced into our sample binaries. We also performed simulations where we corrupted the target addresses kept in memory or registers for different branch types (both calls and returns) in a debugger. In each case, we observed the \monitor detecting and preventing the compromised control flow.   

\subsection{Performance Considerations}
\label{subsec:performance}

\begin{table*}[t]
\begin{center}
  \caption{Microbenchmark results. "\textit{Monitor traps}" shows the number
    of \monitor invocations during the execution of the microbenchmark routine.
    "\textit{Ratio}" shows the ratio of instrumented control-flow transfer
    instructions in relation to other machine code instructions in the main
    program image.}
    \label{tbl:microbenchmarks}
    \begin{tabular}{ | l | c | c | c | c | c |}
    \hline
    \multicolumn{1}{|c|}{\textbf{Program}}          &
    \multicolumn{1}{|c|}{\textbf{Monitor traps}} &
    \multicolumn{1}{|c|}{\textbf{Ratio}} &
    \multicolumn{1}{|c|}{\textbf{Uninstrumented}}  &
    \multicolumn{1}{|c|}{\textbf{Instrumented}}    &
    \multicolumn{1}{|c|}{\textbf{Overhead}}       \\
    \hline
    \tt otp  & 4  & $ \frac{1}{956}$   & 0.53 ms & 0.59 ms & 0.07 ms \ifnotabridged (13\%)  \fi \\
    \tt hmac & 80 & $ \frac{1}{588.4}$ & 0.02 ms & 0.09 ms & 0.07 ms \ifnotabridged (369\%) \fi \\
    \hline
    \end{tabular}
\end{center}
\end{table*}

\begin{table*}[t]
\begin{center}
  \caption{Dhrystone results. The "\textit{One run through Drystone}" column
    shows the average runtime through the Dhrystone benchmark for the
    "\textit{Uninstrumented}" and "\textit{Instrumented}" program versions
    respectively.}
    \label{tbl:dhrystone}
    \begin{tabular}{ | c | c | c | c | c | c |}
    \hline
    \multicolumn{1}{|c|}{\textbf{Monitor}} &
    \multirow{2}{*}{\textbf{Ratio}}        &
    \multicolumn{3}{|c|}{\textbf{One run through Drystone}}  \\
    \multicolumn{1}{|c|}{\textbf{traps}} &
    & 
    \multicolumn{1}{|c|}{\textbf{Uninstrumented}}  &
    \multicolumn{1}{|c|}{\textbf{Instrumented}}    &
    \multicolumn{1}{|c|}{\textbf{Overhead}}       \\
    \hline
    34  & $ \frac{1}{26.4}$ & 0.15 ms & 0.76 ms & 0.61 ms \ifnotabridged (513\%) \fi \\
    \hline
    \end{tabular}
\end{center}
\end{table*}

\ifnotabridged

On Cortex-M processors a \texttt{bl} instruction typically takes a single cycle
+ a pipeline refill ranging between 1 -- 3 cycles to execute depending on the
alignment and width of the target instruction as well as the result of branch
prediction. A subroutine return via a \texttt{bx lr} has a comparable cycle count,
while a return through a \texttt{pop \{\ldots, pc\}} costs an additional cycle
per register to be loaded\footnotemark. As a result of the instrumentation, all
aforementioned instructions have been eliminated from the program image, and
redirected to the \monitor where the additional validations and shadow stack
update takes place. Therefore our instrumentation will increase the runtime
cost (cycle count) of a branch severalfold. Subroutine calls additionally incur
the cost of the branch table lookup to resolve the target of the call. The
runtime cost of the lookup has a logarithmic relation to the size of the branch
table.

\footnotetext{\url{http://infocenter.arm.com/help/index.jsp?topic=/com.arm.doc.ddi0439b/CHDDIGAC.html}}

\fi

The overhead \tzmcfi adds to program execution is dependent on the number of
subroutine calls and returns in the program. We evaluated the impact of \tzmcfi
on performance using microbenchmarks with varying proportions of subroutine calls (and
returns) in relation to other instructions. Our microbenchmarks consisted of an
event-based One-Time Password (OTP) generation algorithm that uses the Corrected
Block Tiny Encryption Algorithm (XXTEA) block cipher algorithm, and a Hash-based
Message Authentication Code (HMAC) implementation using the SHA256 cryptographic
hash function. The size of the branch table was kept constant for each experiment.
Our microbenchmarks contain only direct subroutine calls and all indirect
branches corresponded to effective returns.

We also instrumented the \emph{Dhrystone} 2.1 benchmark
program~\cite{Weicker84} in order to estimate the performance impact on larger
pieces of software. Dhrystone is a synthetic systems programming benchmark used
for processor and compiler performance measurement. It is designed to reflect
actual programming practice in systems programming by modeling the distribution
of different types of high-level language statements, operators, operand types
and locality sourced from contemporary systems programming statistics. In
particular, it attempts to reflect good programming practice by ensuring that
the number of subroutine calls is not too low. Today Dhrystone has largely
been supplanted by more complex benchmarks such as SPEC CPU
bencmarks\footnote{\url{https://www.spec.org/benchmarks.html}} and
CoreMark\footnote{\url{http://www.eembc.org/coremark/about.php}}. The SPEC CPU
benchmarks in particular have been used in prior CFI
literature~\cite{Abadi09,Dang15}. However, the SPEC suite is not practical to
port to MCUs cores. The support library accompanying the Dhrystone benchmark
contains both direct and indirect subroutine calls, and indirect returns. Other
types of indirect branches were not observed in the main program portion of the
samples.
\ifnotabridged
We followed the guidelines in ARM application notes on Dhrystone benchmarking for ARM Cortex Processors~\cite{ARM-Dhrystone}.
\fi

All measurements were performed on an ARM Versatile Express Cortex-M Prototyping
System MPS2+ FPGA configured as a Cortex-M23 processor executing at 25MHz.
Table~\ref{tbl:microbenchmarks} shows the results of the microbenchmarks and
Table~\ref{tbl:dhrystone} shows the result for the Dhrystone benchmarks. According to
the measurements the overhead of \tzmcfi ranges between 13\% -- 513\%.  The
results compare favorably to existing software protection based shadow stack
schemes with reported overheads ranging between 101\% -
4400\%~\cite{Chiueh01,Erlingsson06} (see Section~\ref{sec:related-work}).

\subsection{Memory Considerations}
\label{subsec:codesize}

While layout preserving instrumentation does not add instructions to the program
image, the \monitor and the constructed branch and call target tables and need
to be placed in device memory.
The \monitor only needs to include the logic to handle branch
variants present for a particular program image. For our microbenchmark program
image the \monitor implementation adds a fixed 700 bytes (5.1\%) to the program
image size. The branch table for the microbenchmarks program binary consists of
75 8-byte records, adding 600 bytes (4.3\%) to the program image. Overall the
memory consumption of our microbenchmark program increased by 9.4\%. For out
Dhrystone program image the \monitor adds \monitor 1143 bytes (5.5\%) and the
branch and call target tables 1528 bytes (7.3\%) and 376 bytes (1.7 \%). Overall
the memory consumption of the Dhrystone program increated by 14.5\%). The
numbers for the Dhrystone program include full instrumentation of the support
library.

\section{Extensions}
\label{sec:extensions}

\paragraph{Function-Reuse Attacks.}

The call target validation as presented in Section~\ref{sec:tzmcfi} does address issue of complete function reuse attacks within the main program code. An attacker might perform a pointer substitution where a pointer to one subroutine is exchanged for another subroutine. As both constitute valid call targets, the control-flow transfer would be allowed. Our instrumentation tools allow a human analyst to reduce the set of subroutines that may be subsitituted for each other by limiting the entries to the call target table known to be targeted by indirect subroutine calls, e.g. subrutines used as callback functions. However, as the call target may be computed by the program only at runtime, it is impractical to programatically fully explore all possible execution paths of a program during static analysis and pre-compute a complete CFG. This remains an open problem for any CFI scheme.

\paragraph{Threading}

In our current implementation, the normal state software is limited to using the Main stack. In order to enable CFI for the rudimentary threading supported by Cortex-M processors, the \monitor must be extended to maintain a separate shadow stack for return addresses on the Process call stack. The changes to the \monitor are straightforward as it can consult the \texttt{spsel} register to determine which shadow stack to update.

\paragraph{On-device instrumentation}

The layout-preserving instrumentation approach described in Section~\ref{sec:instrumentation} has properties that make it suitable for performing binary rewriting on-device. Firstly, since it does not affect addresses resolved at link-time, it can be completed in a single pass over the binary image. Secondly, the logic consists of a simple search and replace of branch instruction patterns and branch table construction. While our current implementation relies on an separate tool for rewriting, it is straighforward to implement the needed instrumentation as part of the installation process on-device.   

\paragraph{Binary patching}

Another approach to performing the instrumentation required for CFI is \emph{Binary patching}~\cite{Brown17}. In binary patching, instrumented instructions are replaced with dispatch instructions to trampolines that are placed in unused memory. Compared to \emph{binary rewriting}~\cite{Habibi15}, binary patching does not require adjusting of \emph{all} \texttt{pc}-relative offsets thanks to minimal impact to the program memory layout. However, as explained in Section~\ref{sec:background}, Thumb-2 code has properties that makes binary patching more challenging compared to the instrumentation approach described in Section~\ref{sec:instrumentation}; dispatch instructions used for ARM binary patching are typically 32-bit Thumb-2 \texttt{pc}-relative branches in order to encode a sufficient offset to reach the trampolines. If the instrumented instruction a 16-bit Thumb instruction, the 32-bit dispatch cannot be inserted without affecting the memory layout of the binary. Instead of adjusting all subsequent instructions, both the 16-bit target instruction, and another (16-bit or 32-bit) is moved to the trampoline to make room for the dispatch instruction. If the moved instruction contains a \texttt{pc}-relative operation, it needs to be adjusted accordingly since the new location of the instruction will have a different \texttt{pc} value. Even for a small instruction set such as Thumb, the required logic to perform such adjustments is not in general practical to be implemented as part of the software update mechanism on a resource constrained device. Additionally, as trampolines may contain instructions moved from the instrumentation point, each instrumentation point requires a corresponding trampoline. However, for use cases where on-device instrumentation may not be a concern, a TZ-M protected shadow stack could be utilized with binary patching. This approach would have the advantage of not requiring \monitor logic in the supervisor call handler.

\section{Related Work}
\label{sec:related-work}

Code-reuse attack countermeasures have been a focal topic of
research for the past decade. The most widely used mitigation technique
against this class of attack is \emph{Address Space Layout Randomization}
(ASLR)~\cite{Cohen93,Larsen14})\ifnotabridged, which is deployed by major operating systems today
\footnote{\url{https://www.microsoft.com/security/sir/strategy/default.aspx\#!section\_3\_3}}
\footnote{\url{https://www.kernel.org/doc/Documentation/sysctl/kernel.txt}}
\footnote{\url{https://web.archive.org/web/20120104020714/http://www.apple.com/macosx/whats-new/features.html\#security}}\fi.
ASLR relies on shuffling the executable (and the stack
and heap) base address around in virtual memory, thus requiring an attacker to
successfully guess the location of the target code (or data).
This makes ASLR impractical for constrained devices that lack MMUs and where
memory is a scarce resource.

Dang et.al.~\cite{Dang15} conduct a comprehensive evaluation of shadow stacks schemes in the face of different adversarial models. Dang et al.’s parallel shadow stack~\cite{Dang15} and many traditional shadow stacks~\cite{Giffin02,Giffin04,Nebenzahl06} are based on \emph{unprotected shadow stacks}, e.g., their integrity can be compromised if the shadow stack location is known as they are located in the same address space as the vulnerable application. Shadow stacks protected by canary values~\cite{Erlingsson06,Prasad03} can withstand attack that are limited to sequantial writes, but not arbitrary writes to specific memory addresses. Dang et al. identify only two schemes that operate under an equivalent adversary model as \tzmcfi, in particular with regard to the ability to withstand disclosure of the shadow stacks location; Chiueh and Hsu’s Read-Only RAD~\cite{Chiueh01} and Abadi et al.’s CFI scheme~\cite{Abadi09}. Read-Only RAD incurs a substantial overhead in the order of 1900\% -- 4400\% according to benchmarks by the authors. Abadi et al.’s protected shadow stack achieves a modest overhead between ~5\% -- 55\% (21\% on average). However, it makes use of x86 memory segments, a hardware feature not available on low-end MCUs. In contrast, \tzmcfi provides equivalent security guarantees without requiring hardware features unique to high-end general purpose processors and
compared to previous work on software-only protected shadow stacks, CaRE performs better.

In addition, we consider an even stronger adversary who can exploit interrupt handling to undermine CFI protection; this has been largely ignored in previous CFI works. Prior work, most notably ROPDefender~\cite{Davi11} and PICFI~\cite{Niu15} support \emph{software exception handling}, particularly C++ exceptions. To the best of our knowledge, \tzmcfi is the first scheme to protect \emph{hardware interrupts} initiated by the CPU, a necessity for CFI in bare-metal programs. We make no claim regarding software exceptions, as our system model assumes C programs.

The prevalence of ROP and JOP exploitation techniques in runtime attacks on modern PC platforms has also prompted processor manufacturers to provide hardware support for CFI enforcement. In June 2016, Intel announced its \emph{Control-flow Enforcement Technology}~\cite{Intel-CET} that adds support for shadow call stacks and indirect call validation to the x86/x84 instruction set architecture. Similarly the ARMv8.3-A architecture provides \emph{Pointer Authentication} (PAC)~\cite{ARM-PAC} instructions for ARM application processors that can be leveraged in the implementation of memory corruption countermeasures such as stack protection and CFI. Countermeasures suitable for resource-constrained embedded devices, however, have received far less attention to date. Kumar et al.~\cite{Kumar07} propose a software-hardware co-design for the AVR family of microcontrollers that places control-flow data to a separate \emph{safe-stack} in protected memory. Francillon et al.~\cite{Francillon09} propose a similar hardware architecture in which the safe-stack is accessible only to return and call instructions. AVRAND by Pastrana et al.~\cite{Pastrana16} constitutes a software-based defense against code reuse attacks for AVR devices. HAFIX~\cite{Davi15} is a hardware-based CFI solution for the Intel Siskiyou Peak and SPARC embedded system architectures.

\section{Conclusion}
\label{sec:conclusion}

Security is paramount for the safe and reliable operation of connected IoT devices. It is only a matter of time before the attacks against the IoT device evolve from very simple attacks such as targeting default passwords to advanced exploitation techniques such as code-reuse attacks. 
The introduction of lightweight trust anchors (such as TrustZone-M) to constrained IoT devices will enable the deployment of more advanced security mechanisms on these devices. We show why and how a well understood CFI technique needs to be adapted to low-end IoT devices in order to improve their resilience against advanced attacks. Leveraging hardware assisted security is an important enabler in \tzmcfi, but it also meets other requirements important for practical deployment on small devices, such as interrupt-awareness, layout-preserving instrumentation and the possibility for on-device instrumentation. For small, interrupt-driven devices, the ability to ensure CFI in both interruptible code, as well for the code executing in interrupt contexts is essential.

\section*{Acknowledgments}
\label{sec:acknowledgments}

This work was supported by the German Science Foundation (project S2, CRC 1119 CROSSING), Tekes --- the Finnish Funding Agency for Innovation (CloSer project), and the Intel Collaborative Research Institute for Secure Computing (ICRI-SC).

\ifllncs\bibliographystyle{splncs03}\fi
\ifacmart\bibliographystyle{ACM-Reference-Format}\fi
\bibliography{tzmcfi}

\ifnotabridged
\appendix
\onecolumn

\section{Excerpt of Branch Monitor Code}
\label{appx:sample-code}

\lstdefinestyle{customc}{
  belowcaptionskip=1\baselineskip,
  breaklines=true,
  frame=L,
  xleftmargin=\parindent,
  language=C,
  showstringspaces=false,
  basicstyle=\footnotesize\ttfamily,
  keywordstyle=\bfseries\color{green!40!black},
  commentstyle=\itshape\color{purple!40!black},
  identifierstyle=\color{blue},
  stringstyle=\color{orange},
}

\lstdefinestyle{customasm}{
  belowcaptionskip=1\baselineskip,
  frame=L,
  xleftmargin=\parindent,
  language=[x86masm]Assembler,
  basicstyle=\footnotesize\ttfamily,
  commentstyle=\itshape\color{purple!40!black},
}

\lstset{basicstyle=\tiny,escapechar=@,style=customc,numbers=left}
\begin{lstlisting}[
  label=lst:monitor-excerpt,
]
#define STATE_BIT (0x00000001U)
#define IS_EXC_RETURN(addr) ((unsigned int)(addr) > (unsigned int)0xF0000000)

void SVC_Handler(void)          /* supervisor exception handler entry point         */
{
  __asm(                        /* inline assembler trampoline to branch monitor    */
    "mrs  r0, msp\n"            /* pass pointer to Main stack top as first argument */
    "b    Branch_Monitor_main"  /* invoke main branch monitor routine               */
  )
}

void Branch_Monitor_main(unsigned int* svc_args)
{
  unsigned int svc_number;                     /* svc comment field number          */
  unsigned int stored_lr;                      /* pointer to lr stored on stack     */
  /* Stack frame contains: r0, r1, r2, r3, r12, r14 (lr), pc and xpsr
   * - r0   = svc_args[0] <- offset from top of stack
   * - r1   = svc_args[1]
   * - r2   = svc_args[2]
   * - r3   = svc_args[3]
   * - r12  = svc_args[4]
   * - lr   = svc_args[5]
   * - pc   = svc_args[6]
   * - xpsr = svc_args[7] */
  svc_number = ((char *)svc_args[6])[-2];       /* read comment field from svc instr */

  switch(svc_number)
  {
    case BL_IMM:                                /* branch with link (immediate)      */
      svc_args[5] = svc_args[6] | STATE_BIT;    /* lr = next_instr_addr<31:1> : '1'; */
      Secure_ShdwStk_Push(svc_args[5]);         /* push ret address on shadow stack  */
      svc_args[6] = btbl_bsearch(svc_args[6]);  /* branch table lookup for dest addr */
      goto out;
    case BX_LR:                                 /* branch and exchange (lr)          */
      if (IS_EXC_RETURN(svc_args[5]) {          /* exception return via bx lr        */
        stored_lr = svc_args[5];                /* lr stored in stack context        */
        svc_args[6] = (uint32_t)(&ret_bx_lr) & ~(0x00000001U);
        goto return_through_trampoline;         /* bx lr trampoline                  */
      }                                         /* fast track bybassing trampoline   */
      if (svc_args[5] != Secure_ShdwStk_Pop())  /* validate return address on stack  */
        { abort(); }                            /* halt on return address mismatch   */
      svc_args[6] = svc_args[5] & ~(STATE_BIT); /* pc = lr<31:1> : '0';              */
      goto out;
    case POP_R4_PC:                             /* pop r4, pc                        */
      /* Stack frame additionally contains:
       * - r4 = svc_args[8]
       * - lr = svc_args[9] */
      stored_lr = svc_args[9];                  /* set pointer to stored lr          */
      svc_args[6] = (uint32_t)(&ret_pop_r4_pc) & ~(STATE_BIT);
      goto return_through_trampoline;           /* pop {r4,pc} trampoline            */
    case POP_R4_R5_PC:                          /* pop {r4,r5,pc}                    */
      /* Stack frame additionally contains:
       * - r4 = svc_args[9]
       * - r5 = svc_args[10]
       * - lr = svc_args[11] */
      stored_lr = svc_args[11];                 /* set pointer to stored lr          */
      svc_args[6] = (uint32_t)&ret_pop_r4_r5_pc & ~(STATE_BIT);
      goto return_through_trampoline            /* pop {r4,r5,pc} trampoline         */
      /* ... */                                 /* addl. cases omitted for brewity   */
    return_through_trampoline:
       if (stored_lr == Secure_ShdwStk_Pop())
         { goto out; }                          /* validate return address on stack  */
    default:
        abort();                                /* unrecognized svc number           */
    out:
        return;                                 /* return to trampoline / call dest. */
}
\end{lstlisting}

\fi

\end{document}